\documentclass[twoside,twocolumn,9pt]{article}
\usepackage{extsizes}
\usepackage[super,sort&compress,comma]{natbib} 
\usepackage[version=3]{mhchem}
\usepackage[left=1.5cm, right=1.5cm, top=1.785cm, bottom=2.0cm]{geometry}
\usepackage{balance}
\usepackage{times,mathptmx}
\usepackage{sectsty}
\usepackage{graphicx} 
\usepackage{lastpage}
\usepackage[format=plain,justification=justified,singlelinecheck=false,font={stretch=1.125,small,sf},labelfont=bf,labelsep=space]{caption}
\usepackage{float}
\usepackage{fancyhdr}
\usepackage{fnpos}
\usepackage[english]{babel}
\addto{\captionsenglish}{%
  
}
\usepackage{array}
\usepackage{droidsans}
\usepackage{charter}
\usepackage[T1]{fontenc}
\usepackage[usenames,dvipsnames]{xcolor}
\usepackage{setspace}
\usepackage[compact]{titlesec}
\usepackage{hyperref}
\usepackage{epstopdf}
\definecolor{cream}{RGB}{222,217,201}

\usepackage{booktabs}

\begin{document}

\pagestyle{fancy}
\thispagestyle{plain}


\makeFNbottom
\makeatletter
\renewcommand\LARGE{\@setfontsize\LARGE{15pt}{17}}
\renewcommand\Large{\@setfontsize\Large{12pt}{14}}
\renewcommand\large{\@setfontsize\large{10pt}{12}}
\renewcommand\footnotesize{\@setfontsize\footnotesize{7pt}{10}}
\makeatother

\renewcommand{\thefootnote}{\fnsymbol{footnote}}
\renewcommand\footnoterule{\vspace*{1pt}%
\color{cream}\hrule width 3.5in height 0.4pt \color{black}\vspace*{5pt}} 
\setcounter{secnumdepth}{5}

\makeatletter 
\renewcommand\@biblabel[1]{#1}            
\renewcommand\@makefntext[1]%
{\noindent\makebox[0pt][r]{\@thefnmark\,}#1}
\makeatother 
\renewcommand{\figurename}{\small{Fig.}~}
\sectionfont{\sffamily\Large}
\subsectionfont{\normalsize}
\subsubsectionfont{\bf}
\setstretch{1.125} 
\setlength{\skip\footins}{0.8cm}
\setlength{\footnotesep}{0.25cm}
\setlength{\jot}{10pt}
\titlespacing*{\section}{0pt}{4pt}{4pt}
\titlespacing*{\subsection}{0pt}{15pt}{1pt}

\renewcommand{\headrulewidth}{0pt} 
\renewcommand{\footrulewidth}{0pt}
\setlength{\arrayrulewidth}{1pt}
\setlength{\columnsep}{6.5mm}
\setlength\bibsep{1pt}

\makeatletter 
\newlength{\figrulesep} 
\setlength{\figrulesep}{0.5\textfloatsep} 

\newcommand{\topfigrule}{\vspace*{-1pt}%
\noindent{\color{cream}\rule[-\figrulesep]{\columnwidth}{1.5pt}} }

\newcommand{\botfigrule}{\vspace*{-2pt}%
\noindent{\color{cream}\rule[\figrulesep]{\columnwidth}{1.5pt}} }

\newcommand{\dblfigrule}{\vspace*{-1pt}%
\noindent{\color{cream}\rule[-\figrulesep]{\textwidth}{1.5pt}} }
\newcommand{\dfdx}[2]{\left(\frac{\partial #1}{\partial #2}\right)}

\makeatother

\twocolumn[ 
\begin{@twocolumnfalse}
\vspace{3cm}
\sffamily
\begin{tabular}{m{4.5cm} p{13.5cm} }
\includegraphics{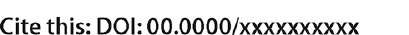} & \noindent\LARGE{\textbf{The Lennard-Jones potential: when (not) to use it.$^\dag$}} \\
\vspace{0.3cm} & \vspace{0.3cm} \\
& \noindent\large
 {Xipeng Wang,\textit{$^{a,b}$} 
 Sim\`{o}n Ram\'irez-Hinestrosa,\textit{$^{b}$}
 Jure Dobnikar, \textit{$^{a,b,c\ddag}$} 
 and Daan Frenkel\textit{$^{a\ast}$}}\\

\includegraphics{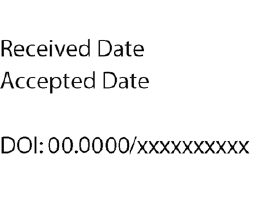} & 
\noindent\normalsize{The Lennard-Jones 12-6 potential (LJ) is arguably the most widely used pair potential in Molecular Simulations.
In fact, it is so popular that the question is rarely asked whether it is fit for purpose. In this paper, we argue that, whilst the LJ potential was designed for noble gases such as argon, it is often used for systems where it is not expected to be particularly realistic. Under those circumstances, the disadvantages of the LJ potential become relevant: most important among these is that in simulations the LJ potential is always modified such that it has a finite range. More seriously, there is by now a whole family of different potentials that are all called {\em Lennard-Jones 12-6}, and that are all different - and that may have very different macroscopic  properties. In this paper, we consider alternatives to the LJ 12-6 potential that could be employed under conditions where the LJ potential is only used as a typical short-ranged potential with attraction. We construct a class of potentials that are, in many respects LJ-like but that are by construction finite ranged, vanishing quadratically at the cut-off distance, and that are designed to be computationally cheap. Below, we present this potential and report numerical data for its thermodynamic and transport properties, for the most important cases (cut-off distance $r_c$=2$\sigma$ (``LJ-like'') and $r_c$=1.2$\sigma$ (a typical ``colloidal'' potential). } \\ 
\end{tabular}
\end{@twocolumnfalse} \vspace{0.6cm} 
]
 

\renewcommand*\rmdefault{bch}\normalfont\upshape
\rmfamily
\section*{}
\vspace{-1cm}


\footnotetext{\textit{$^{a}$~Institute of Physics, Chinese Academy of Sciences, 8 Third South Street, Zhongguancun,
Beijing 100190, China,}}
\footnotetext{\textit{$^{b}$~Department of Chemistry, University of Cambridge, Lensfield Road, CB21EW Cambridge, UK}}
\footnotetext{\textit{$^{c}$~Songshan Lake Materials Laboratory, Dongguan, Guangdong 523808, China}}

\footnotetext{\dag~Electronic Supplementary Information (ESI) available: [details of any supplementary information available should be included here]. See DOI: 10.1039/cXCP00000x/}
\footnotetext{\ddag~jd489@cam.ac.uk}
\footnotetext{$\ast$ \mbox{Corresponding author. df246@cam.ac.uk}}
\section{Background}
One of the most widely used intermolecular potentials in classical many-body simulations, is the so-called Lennard-Jones 12-6 potential 
\begin{equation}
v_{LJ}(r)=4\epsilon\left(\left[{\sigma\over r}\right]^{12} -\left[{\sigma\over r}\right]^6   \right)
\end{equation}
where $\epsilon$ denotes the depth of the attractive well, and $\sigma$ the interparticle distance where the potential changes sign. Lennard-Jones-type  $r^{-n}$-$r^{-m}$ pair potentials were proposed in 1925 by Jones~\cite{jon241} (later Lennard-Jones) to describe the cohesive energy of crystals of noble gases, such as Argon. The now conventional LJ 12-6 form was proposed by Lennard-Jones in 1931~\cite{len311} 
after London had derived that the dispersion  interaction between atoms decays as $r^{-6}$ (at least, in the non-retarded regime)~\cite{lon301}. 
 It later turned out the LJ 12-6 potential is not a particularly good pair potential for Argon~\cite{bar711,mai711}. 
 However, when the first Monte Carlo (MC) simulations of argon were carried out by  Wood and Parker~\cite{woo571} and, subsequently, the first Molecular Dynamics (MD) simulations by Rahman~\cite{rah641}, there was an unexpectedly good agreement between the simulation results and the experimental data for liquid argon. 
 The reason, as  was argued in~\cite{hoe991}  was due to a fortuitous cancellation of errors. 
 However, towards the end of the 20-th century, the LJ 12-6 potential had already achieved an almost unassailable status in classical simulations: it was (and is) used to describe atoms, molecules, coarse-grained models of proteins, and in some cases even larger particles such as nano-colloids. 
 However, for these systems, there is no evidence at all that the LJ 12-6 potential is better than other possible choices. Yet, whenever new simulation techniques are tested, the LJ 12-6 potential is the first model to try. 

However, even if the true Lennard-Jones 12-6 potential would have been a satisfactory all-purpose potential,  there is a practical problem: the Lennard-Jones potential has an infinite range, which would make it less suited for efficient numerical simulations (note that the infinite range was an advantage for (Lennard-)Jones's analytical calculations). 
This problem is normally resolved in practice by truncating the potential at a finite distance $r_c$, e.g. $r_c$=2.5$\sigma$. Unfortunately, not all authors  use the same truncation procedure, and in recent years this confusion has become worse, as the cost of using a larger  (but still finite) cut-off distance has become less prohibitive. In addition, in MD simulations, one should truncate the force, rather than the potential. So actually, the potential is truncated {\em and} shifted. 
Yet even such a force truncation is often not good enough, because a discontinuity in the truncated force creates numerical problems in solving the equations of motion. Therefore, many simulators use a version of the LJ 12-6 potential that is truncated and shifted, {\em and} the force is modified such that it goes to zero continuously at $r_c$. And even that procedure is not unique. So, in the end, there are possibly dozens of different so-called LJ 12-6 potentials in use that are all different, mostly untested for noble gases, and not expected to be particularly good for other molecules either.  All these models have different thermodynamic properties: their critical temperatures may differ by more than 30\% and the liquid-vapour surface tension of these models may differ by much more than that~\cite{bai001}. Finally, this confusion also makes it harder to compare algorithms, if different authors test their favorite algorithm on a different LJ flavor. This problem is well recognized by the community and, in a recent paper, Hafskjold et al~\cite{haf191} have shown that the properties of one particular LJ 12-6 variant (the LJ spline model with $r_c$= 1.74$\sigma$) are quite different from those of the full LJ 12-6 potential.

In summary: the LJ 12-6 potential is anything but a well-defined standard and, in particular for proteins and nano-colloids it is not a good model at all because the range of attraction is too large compared to the effective diameter. There are, of course, LJ n-m style potentials that have been designed to mimic the phase behavior of colloids, but there the range of choices that have been proposed is even larger than for the LJ 12-6 potential.

However, once we give up on the long-ranged dispersion interaction, it is not obvious why a truncated  LJ 12-6 potential should have any special merits that outweigh its disadvantages.  We argue that no such advantage exists. In fact,  even for the quantitative prediction of the properties of noble gases and similarly simple substances, the LJ 12-6 potential is unlikely to remain the model of choice, 
as force-fields based on machine learning applied to ab-initio simulations are increasingly taking over.

What we need for computer experiments (rather than for simulations) is a simple, short-ranged potential that is at least as simple as LJ 12-6, has none of its draw backs and is defined unambiguously.
\section{Constructing a computationally attractive alternative}
If we accept that in most cases of practical interest, the choice of the LJ potential is neither justified by theory nor by experiment, it is logical to ask what requirements should be met by a computationally cheap, generic model pair potential for simple atomistic or molecular systems. Below, we formulate our wish list.
\begin{enumerate}
\item The potential should be repulsive at short distances ($r<\sigma$) and attractive up to a cut-off distance $r_c$ that should not be much larger than $\sigma$. As we explain below, we opt for $r_c=2\sigma$ for atomic systems (which yields a rather LJ-like phase diagram), but for colloidal systems, a smaller value of $r_c$ is preferable (we explore $r_c$=1.2$\sigma $). 
\item A potential to represent simple liquids should have  liquid, vapor and crystalline phases, with a ratio between the critical temperatures and the triple point between $\sim$1.7 (neo-pentane) and $\sim$2.1 (methane). (LJ 12-6 has a ratio $\sim$ 1.91-1.95, and argon $\sim$1.8). In contrast, the ``colloidal'' version of the potential should not exhibit a transition between two thermodynamically stable fluid phases  (``liquid''  and ``vapor''). 
\item At the cut-off, the potential should vanish (at least) quadratically, such that the pair force vanishes continuously at $r_c$.
\item Evaluating the potential should require only few arithmetic operations, and those should be cheap. 
\end{enumerate}
In addition, it is clear that a new pair potential is not very attractive, unless we know its most important thermodynamic and transport properties. In what follows, we will report the dependence of the pressure, energy and free energy on number density $\rho\equiv (N\sigma^3/V)$ and temperature $T\equiv k_BT/\epsilon$. We give the predicted phase diagram, and the liquid-vapor surface tension. And finally, we report the relevant transport properties (diffusivity, viscosity, thermal conductivity) again as a function of $\rho$ and $T$. For the solid phase, we only report the  thermal conductivity.  All data have been fitted to multi-variate polynomials. The raw data are accessible in the Supplementary Information. 
\section{The model}
We wish to construct a simple pair potential that changes from repulsive to attractive at $\sigma$ and that vanishes quadratically at $r_c$. In fact, we also give a more general form  that vanishes as a higher power of $r-r_c$. However, whilst such generalization may have some applications in testing MD algorithms, we do not recommend them. The general form of the potential that we propose is
\begin{equation}\label{eq:potnmrc}
\phi(r)= \epsilon \alpha \left(\left[{\sigma\over r}\right]^{2m} -1 \right)\left(\left[{r_c\over r}\right]^{2m}-1\right)^{2n}\;,
\end{equation}
where $\alpha$ is a coefficient  that ensures that the  depth of the attractive well is $-\epsilon$. $m$ and $n$ are positive integers. We can obtain a simple analytical expression for $\alpha$ and the value of $r$ where the potential reaches its minimum (see Appendix~\ref{app:extrema}). 
 We shall consider the simplest case ($n=m=1$), for which
\begin{eqnarray}\label{eq:n1m1pot}
\phi(r)&\equiv& \epsilon\alpha(1,1;r_c) \left[ \left({\sigma\over r}\right)^2 -1\right]\left[\left({r_c\over r}\right)^2 -1\right]^{2} \mbox{for } r\le r_c\nonumber\\
&=&0 \hspace{4.4 cm}\mbox{for } r> r_c
\end{eqnarray}
with
\begin{equation}
\alpha(1,1;r_c) = {2 \left({r_c\over\sigma}\right)^{2}}\left(\frac{3}{2(\left({r_c\over\sigma}\right)^2-1)}\right)^{3}
\end{equation}
and
\begin{equation}
r_{min}(1,1;r_c)=r_c\left(\frac{3}{1+2\left({r_c\over\sigma}\right)^{2}}\right)^{1/2} \;.
\end{equation}

\subsection{The recommended model}
In what follows,  we use $\sigma$ as our unit of length and $\epsilon$ as our unit of energy.
A particularly simple expression for the pair potential results if $r_c$=2, because $\alpha(1,1;2)$=1. This is our preferred model:  it has a minimum  at $r_{min}\approx$1.155, compared to the LJ 12-6 minimum at $r_{min}(1,1;2)\approx$ 1.1225. The $(1,1:2)$-potential has a relatively short range and is therefore computationally cheap. Moreover, it approaches zero quadratically at $r_c$.   However, for colloidal models, a smaller value of $r_c$ is recommended (we will show results for $r_c=1.2$ with $r_{min}\approx$1.055). 

\begin{figure}[htb]
\centering
\includegraphics[width=\columnwidth]{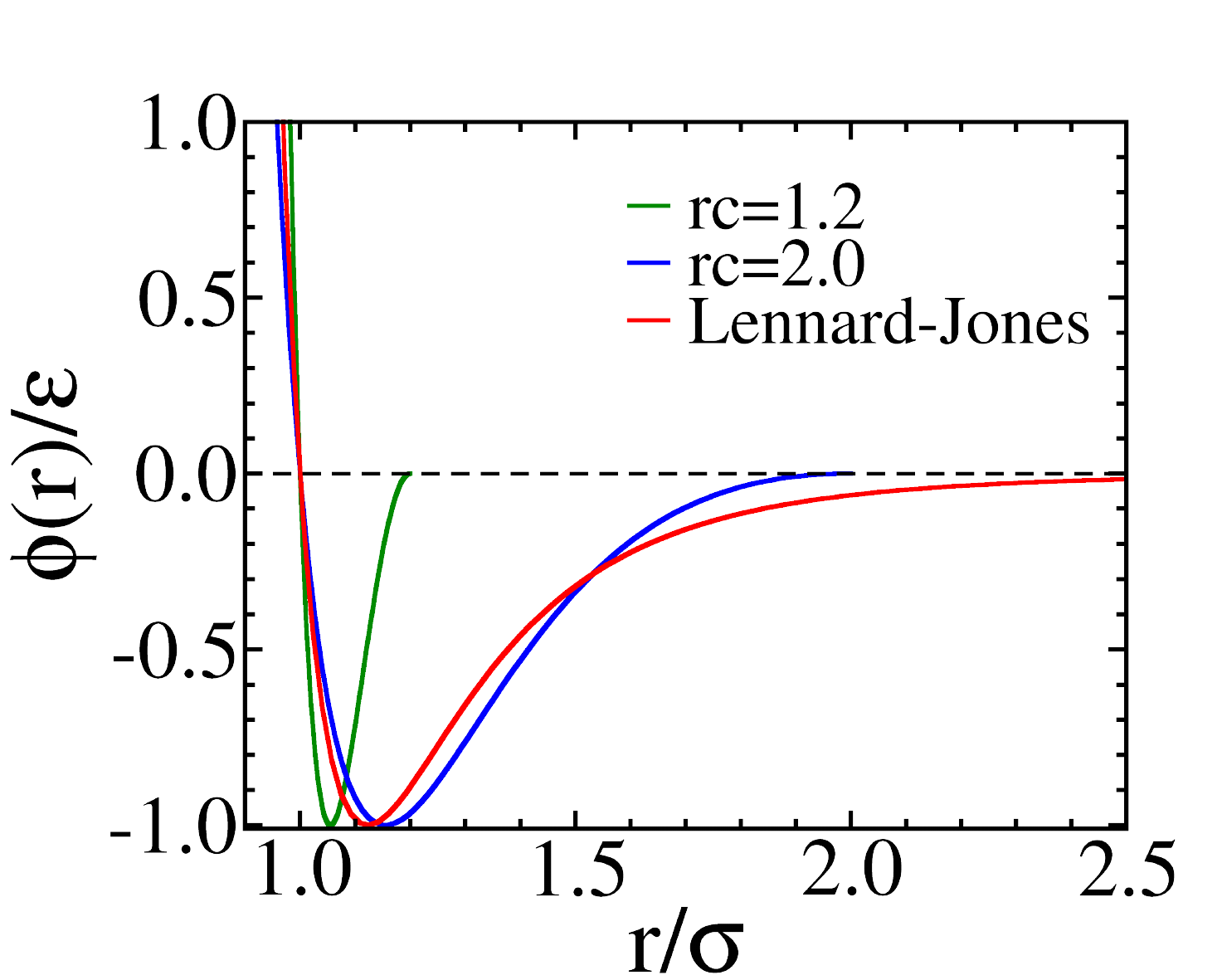}
\captionsetup{width=\columnwidth}
\caption{Comparison of the untruncated Lennard-Jones 12-6 potential (red curve), and the short-ranged potentials described in the text: the ``standard'' form with $m=n=1$ and $r_c$=2.0 is shown as a blue curve. The green curve applies to the the ``colloidal'' form with $m=n=1$ and $r_c$=1.2.}
\label{fig:LJ-GLJ}
\end{figure}
Figure~\ref{fig:LJ-GLJ} shows a comparison of $(1,1,2)$ and $(1,1,1.2)$-versions of simplified potential with the original LJ 12-6 potential.  Of course, there is a wide choice for values of $r_c$, $n$ and $m$, but we argue that, as the advantage of this potential is its simplicity and not its realism for any specific system, there is usually little to be gained by choosing other values of $m$ and $n$. In principle, increasing $n$ would make higher derivatives of the potential continuous at $r_c$, but this comes at a computational cost. The best way to vary the range of the potential is to vary $r_c$. Higher values of $n$ might be interesting when comparing different Molecular Dynamics algorithms. 

Of course, one could also construct a purely repulsive version of these potentials by shifting the potential by $+\epsilon$ and truncating it beyond $r_{min}$. However, there is less need for such a model, as  the repulsive version of the Lennard-Jones 12-6 potential~\cite{wee711} does not suffer from the ambiguities of the attractive LJ potential.  

\section{Properties of the $\mathbf{n=1}$, $\mathbf{m=1}$ model}
One reason why the LJ 12-6 potential is widely used is not its realism, but simply the fact  that by now many observable properties of this model have been computed (see, for instance: equation of state of the  fluid~\cite{tho161,sch181,joh931} and solid~\cite{hoe001,sch181}, transport properties~\cite{lev731,mei041,mei042,mei043} and phase diagram~\cite{han691,tra141,hoe021}).
The  potential given in Eqn.~\ref{eq:n1m1pot} would therefore be of little practical use, if we did not present fairly complete and concise information about its most important equilibrium and transport properties. 

For this reason, we have computed the equation of state, internal energy and free energy of the $n=1$, $m=1$, $r_c=2$  and the $n=1$, $m=1$, $r_c=1.2$ models between low temperatures and a temperature of 1.4 in reduced units for $r_c$=1.2 and 1.6 for $r_c$=2.0, and between low densities and a density of 1.4 (in reduced units). In addition, we have computed the phase diagrams,  the transport properties of the fluid phase (diffusivity, shear viscosity and thermal conductivity), and the liquid-vapor surface tension for the case of  $r_c$=2.0. Finally, we have also computed the thermal conductivity of the crystalline phase. 

All data points can be found in the Supplementary Information (SI). Here we present the coefficients of a multi-variate polynomials fit the describes our simulation data to within the statistical error. To be precise: the fits of the free energy at high $T$ and $\rho$ have a deviation that is slightly larger than the statistical error. We did not try to improve this, as it would make the fit functions more complicated.
\subsection{Results for $\mathbf{r_c}$=2.0}
We first discuss the equation-of-state data. For the fluid, we computed the excess energy and excess pressure, {\em i.e.} the energy and pressure minus the corresponding quantities for an ideal gas at the same temperature and density.

For the solid (we considered face-centered cubic and hexagonal close packed) we also computed the excess energy and pressure, and the excess Helmholtz free energy. 

We performed simulations over a grid of data points for $0 <\rho\le 1.14$ for the liquid, and between $\rho$=0.88 and $\rho$=1.4 for the solid, both over a temperature range between 0.52 and 1.4 (in reduced units). The resolution was 0.02 in both $\rho$ and $T$, although some addition low density points ($\rho <0.02$) where included for the vapor phase. Moreover, we left out a small number of  densities and temperatures from the grid. As all fits turned out to be very smooth, we decided not to fill in these missing points later. All points that turned out to be located in a two-phase region (Liquid-Vapor or Solid-Liquid) where disregarded when performing a fit to the data.  Hence the location of phase boundaries is only based on information about thermodynamically stable state points.

\begin{figure}[htb]
\centering
\includegraphics[width=\columnwidth]{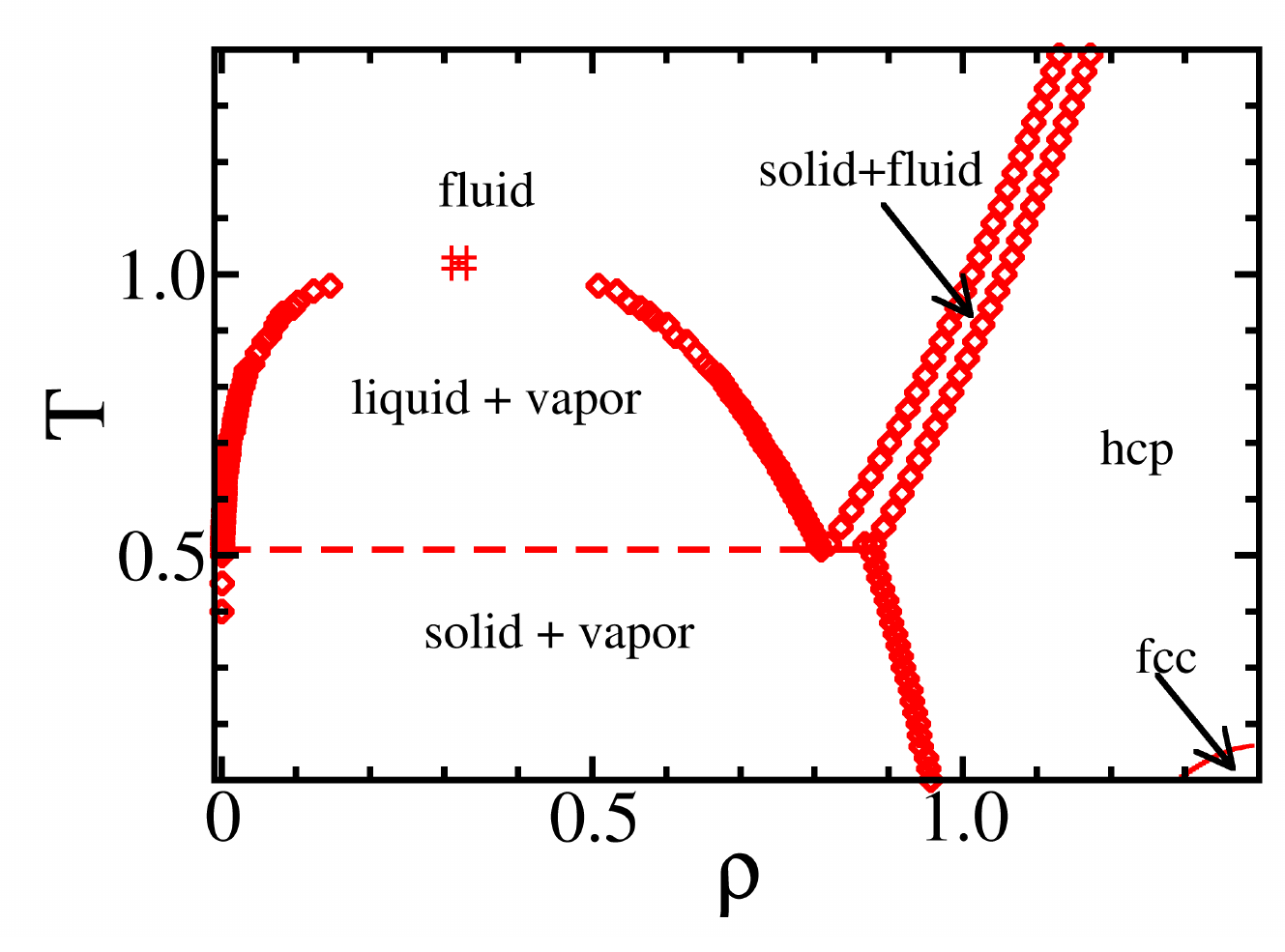}
\captionsetup{width=\columnwidth}
\caption{Computed phase diagram of the potential given by Eqn.~\ref{eq:n1m1pot} for a cut-off distance $r_c$=2.0 (``Lennard-Jones-like''). This phase diagram was computed for a system size of $N$=1000 particles. }
\label{fig:phasediagram2.0}
\end{figure}
\begin{figure}[htb]
\centering
\includegraphics[width=\columnwidth]{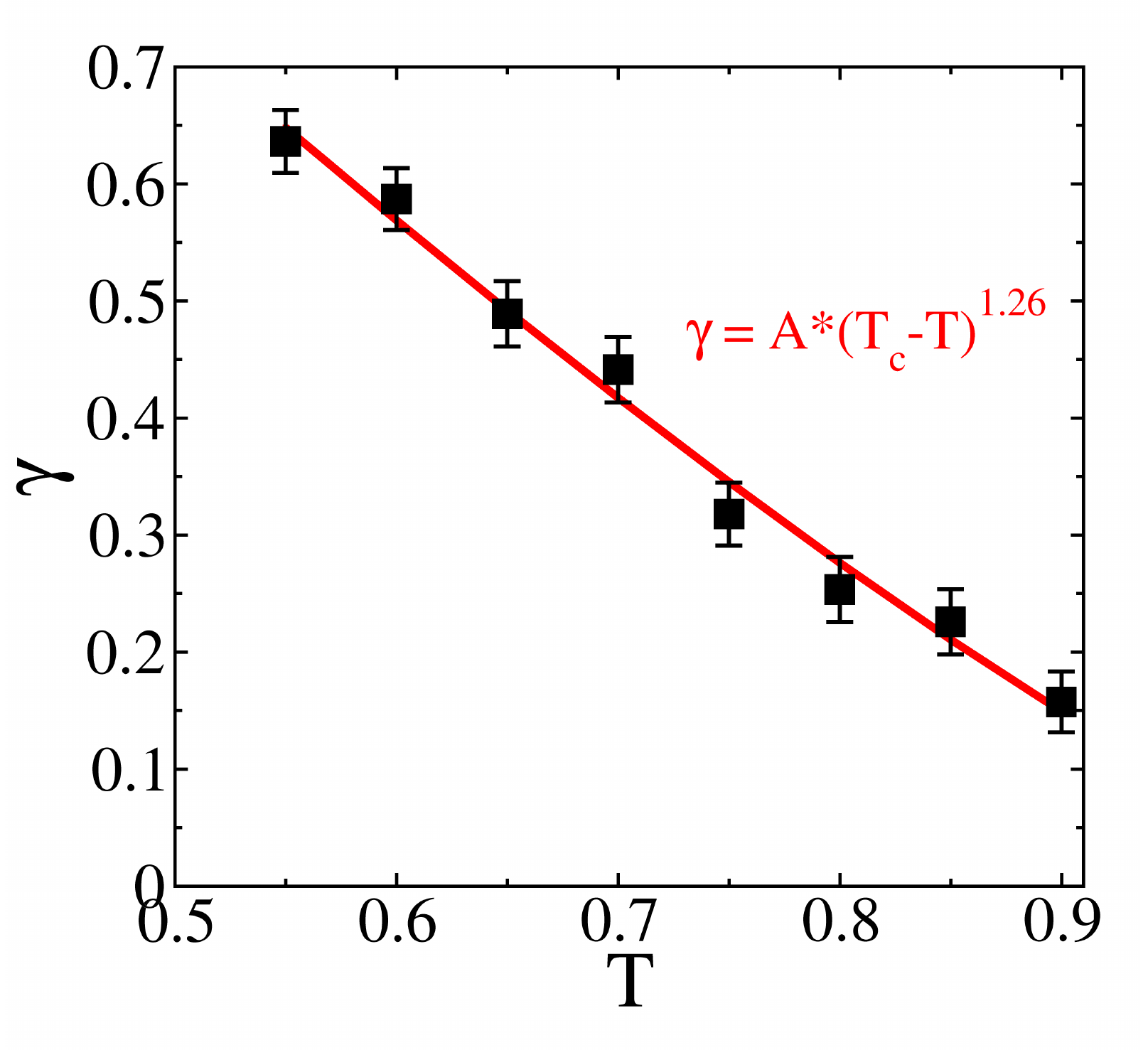}
\captionsetup{width=\columnwidth}
\caption{Temperature dependence of the Liquid-Vapor surface tension of the $r_c=2.0$ model. The curve through the simulation data points was drawn assuming that the surface tension vanishes at the critical point ($T_c=1.04\pm 0.02$) with the 3D Ising critical exponent.}
\label{fig:surface_tension}
\end{figure}

Figure~\ref{fig:phasediagram2.0} shows the phase diagram of the model potential given by Eqn.~\ref{eq:n1m1pot} for a cut-off distance $r_c$=2.0, for a system size of $N$=1000 particles. For this system size, our estimate of the critical temperature is $T_c\approx$ 1.04 with a critical density $\rho_c\approx$0.32. We have not tried to refine the estimate of the critical point using finite-size scaling~\cite{bin811}, as this level of detail is beyond the scope of the present paper.

We estimate the triple point to be located $T_{tr}\approx$ 0.505, where the density of the solid is  $\rho_S\approx$0.883 and that of the coexisting liquid $\rho_L\approx$0.817. We note that the ratio of the critical temperature to that of the triple-point temperature is approximately 2.06, which is higher than the corresponding ratio for argon but lower than that for methane. Figure~\ref{fig:surface_tension} shows the temperature dependence of the Liquid-Vapor surface tension between $T$=0.9$T_c$ and the triple-point temperature. The simulation data agree well with the assumption that the surface tension approaches the value zero at the critical point ($T_c\approx 1.04$) with a power law with the 3D Ising critical exponent. Earlier simulations also found there to be no significant deviations from this scaling form, all the way down to the triple-point temperature. We note that the numerical value of the surface tension is comparable to the values found for various Lennard-Jones variants at the same reduced temperature $T/T_c$.

Note that the slope of the solid-liquid coexistence curves is lower than for the LJ 12-6 system. This is because the repulsive potential in the current model is less steep than in the LJ case ($r^{-6}$ rather than $r^{-12}$). If necessary, the slope could be increased by changing $m$ in the model from $m$=1 to $m\ge$2.

At the triple point, the density of the vapor is extremely low (but can be computed from the knowledge of the chemical potential, which can be computed from the multivariate fit using Eqn.~\ref{eq:mu}). 

For this Lennard-Jones-like model, we expect that the stable solid phase is either face-centered cubic ({\em fcc})  or hexagonal close-packed ({\em hcp}).  In fact, we find both phases. For the $r_c=2.0$ model, the $hcp$ phase appears (very slightly) more stable than $fcc$, except for a small pocket at high densities and low temperatures.  The numerical values of the free energy as a function of density and temperature are given in the SI. We have computed the transport properties and phase boundaries for the $fcc$ crystal phase. Our free-energy calculations showed that, in fact, the $hcp$ phase is slightly more stable for most densities. However, the observable properties of the two phases are so similar that we did not recompute them for the $hcp$ solid. 
\subsection{Results for $r_c$=1.2}
\begin{figure}[t]
\centering
\includegraphics[width=0.9\columnwidth]{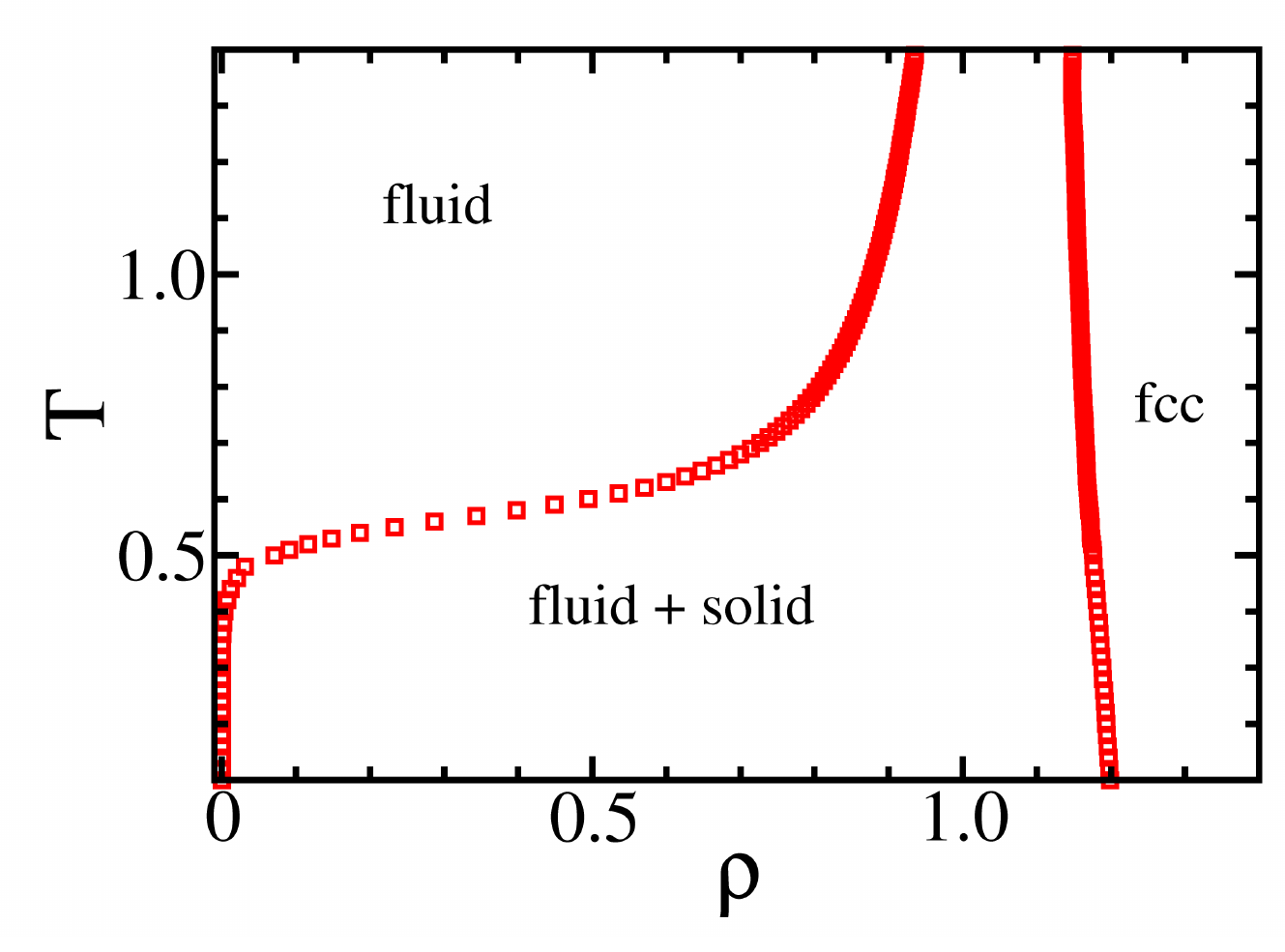}
\captionsetup{width=\columnwidth}
\caption{Computed phase diagram of the potential given by Eqn.~\ref{eq:n1m1pot} for a cut-off distance $r_c$=1.2 (``colloid-like''). This phase diagram was computed for a system size of $N$=1000 particles. }
\label{fig:phasediagram1.2}
\end{figure}

For $r_c=$1.2, the phase diagram shown in Fig.~\ref{fig:phasediagram1.2} does not include a liquid-vapor transition in the stable region. Such behavior is typical for colloidal systems with a relatively short-ranged attractive interaction~\cite{nor201}.
 For the $r_c=1.2$ model,  the $fcc$ crystal phase is more stable than $hcp$, for all densities studied.
Figure~\ref{fig:phasediagram1.2} shows the phase diagram of the model potential given by Eqn.~\ref{eq:n1m1pot} for a cut-off distance $r_c$=1.2, for a system size of $N$=1000 particles. This model corresponds to a typical ``colloidal'' system, which has no liquid-vapor phase transition in the thermodynamically stable region. 
\subsection{Simulation details}
All simulations were carried out using LAMMPS~\cite{pli951}, with a Hamiltonian thermostat, except in the computation of the thermal conductivity, where we used the Nos\'{e}-Hoover thermostat, and compared the results with constant $NVE$ and with the results for a Hamiltonian thermostat: to within the statistical error, we found no difference between the results obtained using these three methods. 
The free-energy calculations for the crystalline solids were carried out in a system of 768 particles, with a periodic box shape that was compatible with both fcc and hcp packing: 12 close-packed planes and 8$\times$ 8 primitive cells in the close-packed planes ($L_x:L_y:L_z$ =$1:\sqrt{3/4}:\sqrt{3/2}$).

 For the direct liquid-vapor coexistence calculations, we simulated a system consisting of $N$=1000 particles in an elongated box with dimensions $L_x:L_y:L_z$=1:1:4. 
 For the thermal conductivity (only for the {\em fcc} solid), we used a box 10$\times$10$\times$9 primitive cells (900 particles). 
 All other simulations were carried out for cubic boxes containing 1000 particles. 
 In all cases, periodic boundary conditions were employed.
 All pressures reported were computed using the virial expression for the stress. 
 \subsection{Transport properties}
 We used Green-Kubo expressions to compute the diffusivity, viscosity and thermal conductivity.  
 We did not attempt to correct for finite-size effects in the transport properties, even though these may be significant~\cite{hey071}. 
 The most important finite-size effect is presumably in the diffusion constant. 
 D\"{u}nweg and Kremer~\cite{dun931} and Yeh and Hummer~\cite{yeh041}, have proposed an expression to correct for this finite size effect for the computation of the diffusivity (at least for a periodically repeated cubic box,as used in this work):
\begin{equation}
\label{eq:D-pbc}
D_\infty = D_{\rm observed}+\frac{k_BT\xi}{6\pi\eta L}  
\end{equation}
where $D_\infty$ is the diffusion constant in an infinite (non-periodic) medium and $D_{\rm observed}$ is the diffusion coefficient observed in a periodic system with box diameter $L$=$(N/\rho)^{1/3}$. 
$\xi$ is a numerical constant withe the value $\xi\approx$2.837297, and $\eta$ denotes the shear viscosity of the fluid. 
The finite-size correction described by Eqn.~\ref{eq:D-pbc} is valid if the fluid behaves as a hydrodynamic continuum on length scales comparable to the system size. However, for highly viscous fluids, reaching this hydrodynamic regime may require very large system sizes.
To our knowledge, there is no numerical evidence for a systematic system-size dependence of the shear viscosity $\eta$~\cite{yeh041}. 
In contrast, finite-size corrections to the thermal conductivity $\kappa$ may be large, in particular for solids (see Chantenne and Barrat who studied finite-size effects in non-equilibrium MD simulations~\cite{cha041}). 
However, there is little information about finite-size effects in the thermal conductivity obtained from Green-Kubo integrals, nor is there much information about such finite-size effect in liquids.
\subsection{Free-energy calculations}
For the free-energy calculations, we used the Einstein-crystal method~\cite{fre841}. 
The maximum spring constant in the Einstein crystal integration was chosen such that the mean-squared  amplitude of the displacement of particles from their lattice sites was the same for the Einstein crystal as for the original crystal. 
To avoid possible problems with the diverging integrand of the thermodynamic integration in the Einstein limit (see {\em e.g.}~\cite{fre021}), we did not start the thermodynamic integration at the Einstein limit, but at a point nearby where the integrand is well behaved. 
To be precise,  writing the potential-energy function  that interpolates between Einstein crystal and real crystal as
\begin{equation}
U(\lambda)=\lambda U_{\rm crystal} +(1-\lambda)U_{\rm Einstein}\;,
\end{equation}
we separated the calculation of the free energy difference into two parts:
\begin{equation}\label{eq:DeltaF}
\beta\Delta F = -\ln<\exp(-\beta \Delta U_{\lambda_{min}})> + \int_{\lambda_{min}}^1
d\lambda \;<\beta\Delta U(\lambda)> \;,
\end{equation}
where 
\begin{equation}
\Delta U(\lambda)\equiv \left\langle  U_{\rm crystal} -U_{\rm Einstein}\right\rangle_\lambda \;.
\end{equation}
For the system size used in the free-energy calculation, we used $\lambda_{\rm min}$=0.0001.  That value could have been further optimised, but optimisation had no noticeable effect on the value or accuracy of the result. The integration in Eqn.~\ref{eq:DeltaF} was carried out using a 10-point Gauss-Legendre quadrature. 

\section{Multi-variate polynomial fits}
All numerical data, can be found in the SI. Here we just present the results of the multi-variate polynomial fits to our numerical data. For convenience and ease of use, we have chosen polynomials as our basic fitting functions. We do not suggest that our choice of the fitting functions is optimal.
\subsection{Thermodynamic properties}

As the data for temperature, pressure and free-energy are related (see below) we used {\em a single} multi-variate fit for excess pressure and energy of the fluid, and a single fit of excess energy, pressure and Helmholtz free energy for the solid. 
The central quantity is the excess Helmholtz free energy $A_{exc}(N,V,T)$, or more precisely $\beta A_{exc}/V$ = $\beta \rho a_{exc}$, where $\beta\equiv 1/k_BT$,  and $a_{exc}\equiv A_{exc}/N$.
We will make use of the following thermodynamic relations:
\begin{equation}
\beta P_{exc}=-\left(\beta \rho a_{exc} -\rho \dfdx{\beta \rho a_{exc}}{\rho}_{T}\right)\;,
\end{equation}
and
\begin{equation}
\rho e_{exc}=  \dfdx{\beta\rho a_{exc}}{\beta}_{\rho}\;.
\end{equation}
To compute the phase diagram, we will also use
\begin{equation}\label{eq:mu}
\beta\mu_{exc} =  \dfdx{\beta\rho a_{exc}}{\rho}_{T}\;.
\end{equation}
Finally, ignoring irrelevant constants, the following relations hold for the ideal gas: $\beta P_{id}$ = $\rho$ and  $\beta \mu^{id}$ = $\ln\rho$. Hence, if we have an expression for $\beta \rho a_{exc}$, we also have expressions for $P$, $e$ and $\mu$.

\begin{table*}
\begin{tabular*}{\textwidth}{@{\extracolsep{\fill}}|l|r|r|r|r|r|r|}
\toprule[\heavyrulewidth]
$\mathbf{n}$&m=$-3$&m=$-2$&m=$-1$&m=$0$&m=$1$&m=$2$\\ 
\midrule
2&0.76902&-3.15391&2.88640&5.17074&-10.4462&0.865638\\ 
3&3.26974&-77.7391&368.594&-723.031&646.553&-219.3018\\ 
4&-8.29352&377.872&-1929.64&3888.37&-3522.43&1219.511\\ 
5&-31.8858&-597.874&3874.72&-8352.09&7764.53&-2720.454\\
6&102.257&349.298&-3783.50&8937.04&-8548.36&3013.089\\
7&-97.1753&-4.10970&1783.38&-4738.69&4672.47&-1647.794\\ 
8&31.1298&-44.7404&-315.575&985.658&-1004.95&353.2964\\ 
\midrule
2&19.1395 &-92.8140&151.915&-99.3440&16.8835& \\ 
3&-21265.4&75376.4&-100238&59293.4&-13166.0& \\ \bottomrule[\heavyrulewidth]
\end{tabular*}
\caption{\label{tab:Fit_ExcessFreeEnergy_Fluid_rc2}
Fit coefficients for the expression for excess free-energy density of the liquid phase and the vapor phase for the model with $r_c$=2.0 (Eqn.~\ref{eq:A_Fluid}). The upper part of the table gives the coefficients that correspond to a fit to the simulation data for all simulated fluid densities at temperatures above $T_c$ and, for supercritical densities below $T_c$. The lower part of the table gives the coefficient for temperatures below $T_c$ and densities below $\rho_c$. In the SI, we give the coefficients with full numerical accuracy. }
\end{table*}

We assume that $\beta \rho a_{exc}$ for phase $\alpha$ can be expanded in a multivariate polynomial. For the liquid we have:
\begin{equation}\label{eq:A_Fluid}
\beta \rho a^{L}_{exc} =\sum_{n=n_{min}}^{n_{max}}\sum_{m=m_{min}}^{m_{max}}  a^{L}_{n,m} \rho^n\beta^m \;.
\end{equation}
However, to reproduce the correct (harmonic) low-temperature behavior of the solid, 
we write 
\begin{equation}\label{eq:A_Solid}
\beta \rho a^{S}_{exc} = \frac{3}{2}\rho\ln\beta +
\sum_{n=n_{min}}^{n_{max}}\sum_{m=m_{min}}^{m_{max}}  a^{S}_{n,m} \rho^n\beta^m \;.
\end{equation}
We then have (for phase $\alpha$=$S$ or $L$):
\begin{equation}
\beta P^\alpha_{exc}(\rho,\beta)=\sum_{n=n_{min}}^{n_{max}}
\sum_{m=m_{min}}^{m_{max}}(n-1)a^\alpha_{n,m}\rho^n\beta^m
\end{equation}
and
\begin{equation}
\rho e^S_{exc} =  \frac{3}{2}\rho T + \sum_{n=n_{min}}^{n_{max}}\sum_{m=m_{min}}^{m_{max}} a^S_{n,m} m\rho^n\beta^{m-1}
\end{equation}
or
\begin{equation}
\rho e^L_{exc} =  \sum_{n=n_{min}}^{n_{max}}\sum_{m=m_{min}}^{m_{max}} a^L_{n,m} m\rho^n\beta^{m-1}
\end{equation}
When fitting the excess pressure and excess energy-density of the fluid/liquid phase, we used  $n_{min}$=2 (because at low densities, the excess pressure scales as $\rho^2$), $n_{max}$=8,$m_{min}$=-3 and $m_{max}$=2.  In the case of $r_c$=2.0, this form was  used for all densities of the fluid at temperature above $T_c$ and for the liquid down to the triple point. Note that all these data points correspond to temperatures above $T_{tr}$.

For the vapor phase ($r_c$=2.0), we used a polynomial with $n_{max}$=6,$m_{min}$=2 and $m_{max}$=2 and $m_{min}$=-3. 

Note that the coefficient of $\rho^2$ follows from the second virial coefficient, that we have computed separately. For the ease of use, it is however, more convenient to treat all fit coefficients on an equal footing (that is: the coefficient of $\rho^2$ was not derived from $B_2(T)$, but was fitted with all the other coefficients. 
\subsection{Transport properties}
We used multi-variate polynomial fits to approximate the diffusivity, viscosity and thermal conductivity of the fluid, and the thermal conductivity of the solid. As the diffusivity $D$ diverges as $1/\rho$ at low densities, we fitted $\rho D$. 
We used the following functional forms: for the diffusivity
\begin{equation}\label{eq:D_rho}
\rho D =\sum_{n=n_{min}}^{n_{max}}\sum_{m=m_{min}}^{m_{max}}  \alpha^{(D)}_{n,m} \rho^n\beta^m \;.
\end{equation}
with $n_{min}$=0,$n_{max}$=3,$m_{min}$=-3 and $m_{max}$=0.
For the viscosity:
\begin{equation}\label{eq:eta_rho}
\eta =\sum_{n=n_{min}}^{n_{max}}\sum_{m=m_{min}}^{m_{max}}  \alpha^{(\eta)}_{n,m} \rho^n\beta^m \;.
\end{equation}
with $n_{min}$=0,$n_{max}$=6,$m_{min}$=0 and $m_{max}$=6.
For the thermal conductivity of the fluid and the solid:
\begin{equation}\label{eq:kappa_rho}
\kappa =\sum_{n=n_{min}}^{n_{max}}\sum_{m=m_{min}}^{m_{max}}  \alpha^{(\eta)}_{n,m} \rho^n\beta^m \;.
\end{equation}
with $n_{min}$=0,$n_{max}$=6,$m_{min}$=0 and $m_{max}$=6.
\begin{table}[h]
\begin{tabular*}{\columnwidth}{@{\extracolsep{\fill}}|l|r|r|r|r|}
\toprule[\heavyrulewidth]

$\mathbf{n}$&m=$-2$&m=$-1$&m=$0$&m=$1$\\
\midrule
0&59.9947&-111.081&23.4348&19.8311 \\ 
1&-254.708&458.412&-85.2926&-88.2034 \\ 
2&427.727&-749.243&124.562&154.892 \\ 
3&-356.028&607.871&-75.9448&-157.340 \\ 
4&147.127&-245.199&21.1821&76.6613 \\ 
5&-24.1761&39.3881&-1.96241&-12.3529 \\ 
\bottomrule[\heavyrulewidth]

\end{tabular*}
\caption{\label{tab:Fit_ExcessFreeEnergy_Solid_rc2}
Fit coefficients for  the expression for excess free-energy density of the {\em fcc} crystal phase  for the model with $r_c$=2.0 (Eqn.~\ref{eq:A_Solid}).  In the SI, we give the coefficients with full numerical accuracy. }
\end{table}

\begin{table}[h]
\begin{tabular*}{\columnwidth}{@{\extracolsep{\fill}}|l|r|r|r|r|}
\toprule[\heavyrulewidth]
$\mathbf{n}$ &m=$-3$&m=$-2$&m=$-1$&m=$0$ \\ 
\midrule
0 & 0.69663 & -2.56961 & 3.31891 & -1.33620 \\
1 & -3.40977 & 12.1701 & -14.5135 & 5.95227 \\
2 & 5.10453 & -17.8776 & 20.7046 & -8.46850 \\
3 & -2.41105 & 8.35025 & -9.54006 & 3.85669 \\
\midrule
0 & & -2.79549 & 5.85772 & -2.94945 \\
1 & & 48.2219 & -97.7224 & 49.7001 \\
2 & & -175.126 & 355.337 & -180.773 \\
\bottomrule[\heavyrulewidth]
\end{tabular*}
\caption{\label{tab:Fit_rhoD_rc2}
Fit coefficients for  the expression for $\rho D$  of the liquid phase and vapor phase for the model with $r_c$=2.0 (Eqn.~\ref{eq:D_rho}). The upper part of the table gives the coefficients that correspond to a fit to the simulation data for all simulated fluid densities at temperatures above $T_c$ and, for supercritical densities below $T_c$. The lower part of the table gives the coefficient for temperatures below $T_c$ and densities below $\rho_c$. In the SI, we give the coefficients with full numerical accuracy. }
\end{table}

\begin{table*}[h]
\begin{tabular*}{\textwidth}{@{\extracolsep{\fill}}|l|r|r|r|r|r|r|r|}
\toprule[\heavyrulewidth]
$\mathbf{n}$&m=$0$&m=$1$&m=$2$&m=$3$&m=$4$&m=$5$&m=$6 $\\ 
\hline
0 &-1025.67 &3674.72 &13.3259 &-17568.4  &31676.6 &-22985.9  &6218.70 \\
1 &3356.81 &-23366.1 &61687.4 &-74998.8  &33867.7  &8135.96 &-8738.63 \\
2 &11489.7 &-51202.6 &71889.4 &-2816.70 &-65814.6  &34706.1  &2083.05 \\
3 &-10159.4 &53558.7 &-115496 &90391.8  &4125.24 &-11895.1 &-11487.9 \\
4 &-1577.01 &50211.2 &-145801 &177803  &-101049  &-5929.13  &27800.4 \\
5 &-43091.4 &147984  &-201402 &147912  &-87827.2 &66770.3 &-31457.5 \\
6 &42832.8 &-191063 &352507  &-348827  &203671  &-75284.1  &16505.3 \\
\midrule
0 &-24.2194 &130.612 &-195.143 &89.0932 & & &\\
1 &-875.221 &1481.66 &-235.286 &-374.664 & & & \\
2 &-2435.26 &5949.15 &-4915.41 &1419.74  & & &\\
3 &15240.2 &-10443.2 &-25016.1 &20194.2  & & &\\

\bottomrule[\heavyrulewidth]
\end{tabular*}
\caption{\label{tab:Fit_eta_rc2}
Fit coefficients for the expression for the shear viscosity $\eta$  of the liquid phase and vapor phase for the model with $r_c$=2.0 (Eqn.~\ref{eq:eta_rho}). The upper part of the table gives the coefficients that correspond to a fit to the simulation data for all simulated fluid densities at temperatures above $T_c$ and, for supercritical densities below $T_c$. The lower part of the table gives the coefficient for temperatures below $T_c$ and densities below $\rho_c$. In the SI, we give the coefficients with full numerical accuracy. }
\end{table*}

\begin{table*}
\begin{tabular*}{\textwidth}{@{\extracolsep{\fill}}|l|r|r|r|r|r|r|r|}
\toprule[\heavyrulewidth]
$\mathbf{n}$&m=0&m=1&m=2&m=3&m=4&m=5&m=6 \\ 
\hline
0  &149.596 &-577.751 &762.037 &-269.597 &-232.326  &210.741 &-42.3878 \\ 
1  &121.498 &-1762.49 &4762.55 &-4701.22  &1831.24 &-467.709  &215.772 \\ 
2  &1267.22 &-1646.78 &-1088.86 &-743.600  &3964.64 &-897.456 &-841.283 \\ 
3 &-2770.85 &6173.25 &-1969.91 &-1254.57 &-591.950 &-883.117  &1299.32 \\ 
4  &65.6882 &-1829.26 &1879.76  &1899.70 &-1720.49 &-1707.09  &1333.25 \\ 
5  &2825.56 &-2287.38 &-6347.60 &6021.19 &-379.015  &3047.14 &-2738.52 \\ 
6 &-1904.17 &3644.62 &-2652.01  &5624.72 &-7964.28  &2738.41  &444.527 \\ 
\midrule
0 &-210.615 & 1042.19 &-1402.49 & 570.548 & & & \\
1 &-54.5982 &-3446.25 & 6110.59 &-2600.56 & & & \\
2 & 2675.78 &-12716.3 & 23099.6 &-13092.7 & & & \\
3 & 55913.5 &-128202 & 78195.6 &-5837.50 & & & \\

\bottomrule[\heavyrulewidth]
\end{tabular*}
\caption{\label{tab:Fit_kappa_rc2}
Fit coefficients for the expression for $\kappa$  of the liquid phase and vapor phase for the model with $r_c$=2.0 (Eqn.~\ref{eq:kappa_rho}). The upper part of the table gives the coefficients that correspond to a fit to the simulation data for all simulated fluid densities at temperatures above $T_c$ and, for supercritical densities below $T_c$. The lower part of the table gives the coefficient for temperatures below $T_c$ and densities below $\rho_c$. In the SI, we give the coefficients with full numerical accuracy. }
\end{table*}

\begin{table*}[h!]
\begin{tabular*}{\textwidth}{@{\extracolsep{\fill}}|l|r|r|r|r|r|r|r|}
\toprule[\heavyrulewidth]
 $\mathbf{n}$&m=0&m=1&m=2&m=3&m=4&m=5&m=6 \\ 
 \hline
0 &  31380.5  & -54228.7   &17548.3  &-1481.55  &-3480.41   &3635.92  &-611.898 \\
1  &-41206.1   & 51105.3  & 22036.7  &-2702.04   &1864.24  &-6366.19   &510.533 \\
2  & 93.1931   & 10395.4  &-34834.1  &-10094.2   &3169.27  & 2109.39   &1711.08 \\
3  & 10319.3   &-4986.17  &-12617.4   &3328.08  & 4016.16  & 2709.28  &-2490.02 \\
4  & 3854.17  & -778.979  & 11250.5   &8097.52  &-67.0914  &-1984.86  &-429.411 \\
5  &-3438.01   &-9826.25   &2586.75   &3108.49  &-6384.14  &-2068.63   &2338.61 \\
6   &225.031   & 4188.17  &-4.36.24  &-3944.54  & 2129.02   &1781.09  &-1023.35  \\
\bottomrule[\heavyrulewidth]
\end{tabular*}
\caption{\label{tab:Fit_kappa_rc20_S}
Fit coefficients for the expression for $\kappa$  of the {\em fcc} crystal phase for the model with $r_c$=2.0 (Eqn.~\ref{eq:kappa_rho}).  In the SI, we give the coefficients with full numerical accuracy. }
\end{table*}

\subsection{Fits for $r_c$= 2.0}
\subsubsection{Thermodynamic Properties}
Table~\ref{tab:Fit_ExcessFreeEnergy_Fluid_rc2} summarizes the fit coefficients for Eqn.~\ref{eq:A_Fluid} for the liquid phase, and separately for the vapor phase, for the model with $r_c$=2.0.

Table~\ref{tab:Fit_ExcessFreeEnergy_Solid_rc2} summarizes the fit coefficients for Eqn.~\ref{eq:A_Solid} for the crystalline {\em fcc} phase for the model with $r_c$=2.0.
\begin{table*}[h]
\begin{tabular*}{\textwidth}{@{\extracolsep{\fill}}|l|r|r|r|r|r|r|}
\toprule[\heavyrulewidth]
 $\mathbf{n}$&m=-3&m=-2&m=-1&m=0&m=1&m=2  \\ 
\hline

2&-1.10009&6.27642&-13.5539&15.5901&-7.05497&0.63096 \\ 
3&4.79833&-31.7208&76.0381&-84.1657&46.4876&-10.4596 \\ 
4&-0.47723&19.8875&-62.2031&78.4142&-55.7024&18.4785 \\ 
5&-7.60729&39.2723&-131.249&205.539&-110.462&9.10103 \\ 
6&-22.4105&65.7709&66.3596&-315.674&244.739&-43.4275 \\ 
7&56.4536&-228.002&243.158&30.6737&-133.904&33.7645 \\ 
8&-29.5309&127.279&-174.625&68.3272&18.5640&-9.79081 \\ 
\bottomrule[\heavyrulewidth]
\end{tabular*}
\caption{\label{tab:Fit_ExcessFreeEnergy_Fluid_rc12}
Fit coefficients for the expression for excess free-energy density of the fluid phase for the model with $r_c$=1.2 (Eqn.~\ref{eq:A_Fluid}). In the SI, we give the coefficients with full numerical accuracy. }
\end{table*}

\subsubsection{Transport Properties}

In Tables ~\ref{tab:Fit_rhoD_rc2}-\ref{tab:Fit_kappa_rc2} we summarize the fitting coefficients for the diffusivity, more precisely the product $\rho D$ (Eqn.~\ref{eq:D_rho}); the shear viscosity (Eqn.~\ref{eq:eta_rho}) and the thermal conductivity (Eqn.~\ref{eq:kappa_rho}).
All of these transport coefficients were computed for the fluid phase and we have not attempted to include the data for the low-density gas ($\rho$< 0.1), as in this regime, the Green-Kubo integrals are relatively noisy and the transport coefficients can better be calculated from the Boltzmann equation, using the approach of Chapman and Cowling~\cite{cha521}

Table~\ref{tab:Fit_kappa_rc20_S} summarizes the fit coefficients for Eqn.~\ref{eq:kappa_rho} for the thermal conductivity of the {\em fcc} solid phase for the model with $r_c$=2.0.

\begin{table}[h]
\begin{tabular*}{\columnwidth}{@{\extracolsep{\fill}}|l|r|r|r|r|}
\toprule[\heavyrulewidth]
 $\mathbf{n}$&m=-2&m=-1&m=0&m=1\\ 
\hline
0&1585.85&-2137.72&-1114.08&-0.10780 \\ 
1&-6114.51&8239.19&4047.61&-195.525 \\ 
2&9422.37&-12692.5&-5907.11&682.374 \\ 
3&-7253.89&9769.46&4342.62&-807.773 \\ 
4&2789.93&-3757.12&-1600.17&364.210 \\ 
5&-428.864&577.560&236.283&-46.0042 \\ 
\bottomrule[\heavyrulewidth]
\end{tabular*}
\caption{\label{tab:Fit_ExcessFreeEnergy_Solid_rc12}
Fit coefficients for the expression for excess free-energy density of the {\em fcc} crystal phase  for the model with $r_c$=1.2 (Eqn.~\ref{eq:A_Solid}).  In the SI, we give the coefficients with full numerical accuracy. }
\end{table}
\subsection{Fits for $r_c$=1.2}
\subsubsection{Thermodynamic Properties}
Table~\ref{tab:Fit_ExcessFreeEnergy_Fluid_rc12} summarizes the fit coefficients for Eqn.~\ref{eq:A_Fluid} for the fluid phase, for the model with $r_c$=1.2.

Table~\ref{tab:Fit_ExcessFreeEnergy_Solid_rc12} summarizes the fit coefficients for Eqn.~\ref{eq:A_Solid} for the crystalline {\em fcc} phase for the model with $r_c$=1.2.


\begin{table}[h]
\begin{tabular*}{\columnwidth}{@{\extracolsep{\fill}}|l|r|r|r|r|}
\toprule[\heavyrulewidth]
 $\mathbf{n}$ &m=-3&m=-2&m=-1&m=0 \\
 \hline
0 & 0.04176 &-0.15423 & 0.30690 &-0.03915\\ 
1 &-0.09000 & 0.28129 &-0.36580 &0.25672\\ 
2 & 0.41788 &-1.36771 & 1.48079 &-0.90280\\ 
3 &-0.47069 & 1.57829 &-1.76868  &0.80454\\ 
\bottomrule[\heavyrulewidth]
\end{tabular*}
\caption{\label{tab:Fit_rhoD_rc12}
Fit coefficients for  the expression for $\rho D$  of the fluid phase for the model with $r_c$=1.2 (Eqn.~\ref{eq:D_rho}).  In the SI, we give the coefficients with full numerical accuracy. }
\end{table}

\begin{table*}[h]
\begin{tabular*}{\textwidth}{@{\extracolsep{\fill}}|l|r|r|r|r|r|r|r|}
 \toprule[\heavyrulewidth]
 $\mathbf{n}$&m=0&m=1&m=2&m=3&m=4&m=5&m=6\\ 
\hline
0 &-21.0627 & 73.8838 &-101.523 & 76.3315 &-35.9647 & 9.97495 &-1.10172 \\
1  &304.977 &-949.825 & 1118.91 &-742.064  &358.002 &-109.781 & 11.2826 \\
2 &-1514.92 & 4454.07 &-5269.71 & 4512.62 &-3153.32 & 1166.78 &-126.363 \\
3  &2031.52 &-3468.57 & 1668.04 &-5065.03 & 7428.87 &-3017.70 & 160.901 \\
4  &920.307 &-9087.33 & 15730.5 &-2095.78 &-7314.55 & 1478.52 & 894.480 \\
5 &-897.404 & 2981.78 & 3051.34 &-16181.4 & 10282.7 & 2798.33 &-2560.19 \\
6  &24.0482 & 611.894 &-1780.61 &-1038.12 & 8783.96 &-9202.61 & 2810.34 \\

\bottomrule[\heavyrulewidth]
\end{tabular*}
\caption{\label{tab:Fit_eta_rc12}
Fit coefficients for  the expression for $\eta$  of the fluid phase for the model with $r_c$=1.2 (Eqn.~\ref{eq:eta_rho}).  In the SI, we give the coefficients with full numerical accuracy. }
\end{table*}

\subsubsection{Transport Properties}
In Tables ~\ref{tab:Fit_rhoD_rc12}-\ref{tab:Fit_kappa_rc12_F} we summarize the fitting coefficients for the diffusivity, more precisely the product $\rho D$ (Eqn.~\ref{eq:D_rho}); the shear viscosity (Eqn.~\ref{eq:eta_rho}) and the thermal conductivity (Eqn.~\ref{eq:kappa_rho}).
All of these transport coefficients were computed for the fluid phase and we have not attempted to include the data for the low-density gas ($\rho$< 0.1), as in this regime, the Green-Kubo integrals are relatively noisy and the transport coefficients can better be calculated from the Boltzmann equation, using the approach of Chapman and Cowling~\cite{cha521}

\begin{table*}[h]
\begin{tabular*}{\textwidth}{@{\extracolsep{\fill}}|l|r|r|r|r|r|r|r|}
\toprule[\heavyrulewidth]
 $\mathbf{n}$ &m=0&m=1&m=2&m=3&m=4&m=5&m=6 \\ 
 \hline
0 & -2.03719 &  114.606 & -433.051 &  672.146 & -528.419 &  210.761 & -34.1161 \\
1 & -459.380 &  592.988 &  2059.10 & -5723.11  & 5714.97 & -2669.76 &  489.759 \\
2 &  4792.68 & -15213.3 &  14970.1 &  1976.83 & -14492.9 &  10377.1 & -2424.31 \\
3 & -8414.27 &  30113.4 & -43241.1 &  21722.6 &  14714.3 & -21369.3 &  6537.43 \\ 
4 & -12862.1 &  44845.5 & -40351.8 &  2485.73 & -9523.57  & 25508.3 & -10224.1 \\
5 &  31717.7 & -98686.9 &  69562.6 &  46534.2 & -58861.8  & 3308.63  & 6557.54 \\
6 & -16527.1 &  48302.2 & -27515.2 & -33077.9 &  34935.2 & -2969.87 & -3195.33 \\
\bottomrule[\heavyrulewidth]
\end{tabular*}
\caption{\label{tab:Fit_kappa_rc12_F}
Fit coefficients for the expression for $\kappa$  of the fluid phase for the model with $r_c$=1.2 (Eqn.~\ref{eq:kappa_rho}).  In the SI, we give the coefficients with full numerical accuracy. }
\end{table*}

Table~\ref{tab:Fit_kappa_rc12_S} summarizes the fit coefficients for Eqn.~\ref{eq:kappa_rho} for the thermal conductivity of the {\em fcc} solid phase for the model with $r_c$=1.2.

\begin{table*}[h]
\begin{tabular*}{\textwidth}{@{\extracolsep{\fill}}|l|r|r|r|r|r|r|r|}
\toprule[\heavyrulewidth]
 $\mathbf{n}$ &m=0&m=1&m=2&m=3&m=4&m=5&m=6 \\ 
 \hline
0&-84237.9&214416&-198388&39887.8&-1683.09&22660.5&-3314.31 \\ 1&126293&-179489&231537&-15974.1&-42312.7&-12879.4&3719.35 \\ 2&-79926.1&-103018&-50331.8&69792.9&-23547.0&4661.70&-8736.66 \\ 
3&57608.1&25353.5&-70553.9&26570.2&-10978.8&4057.21&7647.01 \\ 
4&1760.96&137339&-34058.8&-6416.11&33321.7&1897.50&2704.36 \\ 
5&-62117.8&-16507.1&18451.1&-32836.2&-4996.54&-5931.71&-6129.40 \\ 
6&32370.3&-42227.5&38845.9&-16073.6&11788.1&-2072.94&2478.03 \\ 
\bottomrule[\heavyrulewidth]
\end{tabular*}
\caption{\label{tab:Fit_kappa_rc12_S}
Fit coefficients for the expression for $\kappa$  of the {\em fcc} crystal phase for the model with $r_c$=1.2 (Eqn.~\ref{eq:kappa_rho}).  In the SI, we give the coefficients with full numerical accuracy. }
\end{table*}
\section{Conclusion}
In this paper we have presented a simple model pair-potential that can be used in numerical studies of systems of particles with a short-ranged attraction. The model potential that we use vanishes quadratically at a cut-off distance $r_c$. 
Our main reason for proposing this simple model is that, in practice, the widely used LJ 12-6 model is implemented differently (truncated, shifted, interpolated etc.) by different authors. In contrast, the current model is uniquely defined once $r_c$ is specified.

We report a fairly complete set of thermodynamic and transport properties for the cases $r_c$=2.0 (``atomic liquid'') and $r_c$=1.2 (``colloidal suspension'').
The raw simulation data are available online. 

We stress that the models that we present are not designed to model any specific substance. Rather, they represent generic models. However, in many cases the ubiquitous LJ 12-6 potential is used in exactly the same way. 

We note that the present model can be easily be extended to describe purely repulsive interactions. However, as the Weeks-Chandler-Andersen representation of the LJ 12-6 potential is not ambiguous, there is less need for a ``repulsive'' version of the current potential. 

Finally, we note that the present model shows LJ-like behavior for a cut-off radius $r_c=2.0$, which should make it cheaper than most LJ 12-6 models that tend to have larger cut-off radii and therefore have more interacting neighbors. 
\clearpage

\appendix
\section{Minimum of generalized potential}\label{app:extrema}
Here we derive the expressions for the location and depth of the minimum of the generalized $n,m,r_c$ potential.
Note that in the main text, we use $n=1$, $m$=1 and $r_c=$2.0 and 1.2.

To locate the minimum of the potential of the type given in Eqn.~\ref{eq:potnmrc}, we define an auxiliary quantity $u$ as
\begin{equation}
u\equiv \left[{\sigma\over r}\right]^{2m}
\end{equation}
and 
\begin{equation}
u_c\equiv \left[{\sigma\over r_c}\right]^{2m}\;.
\end{equation}
As before, we use $\sigma$ as our unit of length and $\epsilon$ as our unit of energy.
Then
\begin{equation}
\phi(u)=\alpha \left(u -1 \right)\left({u\over u_c} -1\right)^{2n}\;,
\end{equation}
The condition for the extremum (minimum) is:
\begin{equation}
0=\dfdx{\phi(u)}{u}=\alpha\left[\left({u\over u_c} -1\right)^{2n} +{2n\over u_c}\left(u-1\right)\left({u\over u_c} -1\right)^{2n-1}\right]\;,
\end{equation}
which implies that at $u_{min}$
\begin{equation}
\left({u_{min}\over u_c} -1\right)=-{2n\over u_c}\left(u_{min}-1\right)\;,
\end{equation}
and hence
\begin{equation}
u_{min}(1+2n)=u_c+2n
\end{equation}
or
\begin{equation}\label{eq:umin}
u_{min}=\frac{u_c+2n}{1+2n} \;.
\end{equation}
Eqn~\ref{eq:umin} implies that
\begin{eqnarray}
r_{min}(n,m;r_c)&=&\left(\frac{1}{u_{min}}\right)^{1/2m}\nonumber\\
&=&r_c\left(\frac{1+2n}{1+2nr_c^{2m}}\right)^{1/2m}
\end{eqnarray}
and
\begin{equation}
\alpha(n,m;r_c) = {2n r_c^{2m}}\left(\frac{1+2n}{2n(r_c^{2m}-1)}\right)^{2n+1}
\end{equation}
\section*{Acknowledgements}
DF acknowledges more than 40 years of stimulating discussions with Michiel Sprik, who does not take anything for granted. We gratefully acknowledge discussions with J\"urgen Horbach.
\bibliography{GLJ_refs}
\bibliographystyle{rsc} 
\end{document}